\documentclass{ws-ijmpe}

\begin{document}

\markboth{H.-T. Ding, A. Dainese, Z. Conesa del Valle and D.
Zhou}{Studying the energy loss of heavy quarks via single muon
production in PbPb collisions at $\sqrt{s_\mathrm{NN}}$=5.5~TeV}

\catchline{}{}{}{}{}

\title{STUDYING THE ENERGY LOSS OF HEAVY QUARKS VIA SINGLE MUON PRODUCTION IN PbPb COLLISIONS \\
AT $\sqrt{s_\mathrm{NN}}$=5.5~TeV}

\author{\footnotesize HENG-TONG DING$^{1,2,\dagger}$, ANDREA DAINESE$^{3,\dagger\dagger}$,
ZAIDA CONESA DEL VALLE$^{4,\dagger\dagger\dagger}$ and DAICUI
ZHOU$^{1,\dagger\dagger\dagger\dagger}$ }

\address{$^1$Institute of Particle Physics, Central China Normal University,Wuhan 430079,
China \\
$^2$Centro Enrico Fermi, Via Panisperna 89 A, I-00184 Rome,
 Italy \\
$^3$INFN - Laboratori Nazionali di Legnaro, viale dell'Universit\`a
2, 35020 Legnaro (Padova), Italy \\
$^4$Subatech (CNRS/IN2P3 - Ecole des Mines - Universit\'e de
Nantes) Nantes, France \\
$^{\dagger}$dinght@iopp.ccnu.edu.cn \\
$^{\dagger\dagger}$andrea.dainese@lnl.infn.it \\
$^{\dagger\dagger\dagger}$zconesa@subatech.in2p3.fr \\
$^{\dagger\dagger\dagger\dagger}$dczhou@mail.ccnu.edu.cn}

\maketitle

\begin{history}
\received{(received date)}
\revised{(revised date)}
\end{history}

\begin{abstract}
The effects of heavy quarks energy loss on the transverse momentum
spectra of single muons are studied. The energy loss of heavy
quarks when traversing the medium formed in PbPb collisions at
$\sqrt{s_\mathrm{NN}}$=5.5 TeV is calculated by implementing the
collision geometry and the quenching weights. The medium density
is assumed to decrease at forward pseudo-rapidity and to be
proportional with pseudo-rapidity multiplicity $dN/d\eta$. Muons
from W decays can be used as a medium-blind reference to quantify
the effect of heavy quarks energy loss on the single muon
production.
\end{abstract}

\section{Introduction}
\label{Introuduction}
Heavy quarks are sensitive probes of the medium produced in the
collision as they are produced in the early times and may lose
energy when traversing the medium. At the Relativistic Heavy Ion
Collider (RHIC), a significant suppression of electrons from heavy
quarks at mid-rapidity has been observed in central AuAu
collisions, indicating substantial energy loss of heavy
quarks\cite{PHENIX,STAR}. At the Large Hadron Collider (LHC), the
c.m. energy of PbPb collision is 5.5$~$TeV per nucleon-nucleon
collision, about 30 times larger than that at RHIC and it is
expected that heavy flavors will be abundantly produced, offering
a good opportunity to characterize the medium formed in the
collision. Heavy quark decays are expected to dominate the
lepton-pair continuum up to the mass of the
$Z^{0}$~\cite{ALICEPPR1,Kniehl,Lokhtin,MuonAlice1,MuonAlice2} and
the transverse momentum spectra of single muons up to 30
GeV$/c$~\cite{Zaida}. The shape of the spectra could be altered by
the energy loss of heavy quarks\cite{Lokhtin,LinHVQ,DimuonEloss}.
At the same time, this high amount of energy in the center of mass
enables the possibility to create the W bosons in the initial hard
scattering. As the muons from W decay are not effected by the
medium formed in the collision, it is interesting to see how much
the energy loss of heavy quarks affects on the muon production by
using the muons from W as a reference.

In the present work, we study the effect of heavy quarks energy
loss on the transverse momentum spectra of single muons in PbPb
collisions at $\sqrt{s_\mathrm{NN}}=5.5~\mathrm{TeV}$ in the
acceptance of ALICE MUON Spectrometer\cite{ALICEPPR1}. First, by
introducing the nuclear shadowing effect and $k_{t}$ broadening in
the HVQMNR program\cite{HVQMNR}, which provides a fully exclusive
partonic differential cross section calculation including all the
relevant partonic subprocesses, we calculate the kinematic
distributions of heavy quarks per nucleon-nucleon collision in
PbPb collisions at $\sqrt{s_\mathrm{NN}}=5.5~\mathrm{TeV}$. Then,
to sample the energy loss of heavy quarks, we use the collision
geometry and the BDMPS quenching weights for massive
case\cite{MassiveQuenching} with a scaling of the transport
coefficient as a function of the pseudo-rapidity. Finally we give
the results of the transverse momentum spectra of single muons
within the acceptance of ALICE MUON Spectrometer.

\section{Production of Heavy Quarks}
\label{Production} In the HVQMNR program, the heavy quark hard
production cross section is calculated in pQCD according to the
factorization theorem:
\begin{equation}
\mathrm{d}\sigma_{ab}(\sqrt{s})=\sum_{i,j}\int\mathrm{d}x_{i}\mathrm{d}x_{j}
\mathrm{d}\hat{\sigma}_{ij}(\hat{s},m^{2},\mu_{F}^{2},\mu_{R}^{2})f_{i}^{a}(x_{i},\mu_{F}^{2})
f_{j}^{b}(x_{j},\mu_{F}^{2}),
\end{equation}
where $\mu_{F}$ is the fractorization scale, $f_{i}^{a}(x_{i})$
and $f_{j}^{b}(x_{j})$ are the parton distribution functions, the
differential probabilities for the partons $i$ and $j$ to carry
momentum fractions $x_{i}$ and $x_{j}$ of their respective protons
$a$ and $b$. The total partonic cross section
$\mathrm{d}\hat{\sigma}_{ij}$ of the process $ij\rightarrow
Q\bar{Q}X$ at parton-parton c.m. energy $\hat{s}=x_{i}x_{j}s$,
where $\sqrt{s}$ is the c.m. energy of the pp collision, is
calculated as a perturbation series in $\alpha_{s}(\mu_{R})$,
where the strong coupling constant is evaluated at the
renormalization scale $\mu_{R}$.

 In the present calculation, CTEQ 4M parton
distributions\cite{CTEQ4M} are used and nuclear shadowing is taken
into account by using the EKS98 parameterization\cite{EKS98}. The
parameters of HVQMNR program are set as: for charm, $m_c=1.2$ GeV
and $\mu_F=\mu_R=2\mu_0$, for beauty, $m_b=4.75$ GeV and
$\mu_F=\mu_R=\mu_0$, where $\mu_0\equiv \sqrt{m_Q^2+(p_{t,Q}^2 +
p_{t,\bar{Q}}^2)/2}$. Calculations are performed up to next
leading order (NLO) with the HVQMNR program.

\section{Energy Loss of Heavy Quarks}
\label{Eloss} For modelling the energy loss of heavy quarks, we
used the quenching weights in the multiple soft scattering
approximation, which were derived in Ref.\cite{MassiveQuenching}
in the framework of the BDMPS formalism\cite{BDMPS}.
The energy loss $\Delta E$ is sampled by implementing the
collision geometry and the quenching weights as explained in
Ref.\cite{PQM}.

However, the transport coefficient $\hat{q}$
mentioned in the Ref.\cite{PQM} is only applied for the central
rapidity region. The decreased medium density in the forward
rapidity region should be taken into account. We assume the
scaling
\begin{equation}
\hat{q}(\eta^{\ast})~=~\hat{q}(0)\cdot\frac{\mathrm{d}N_{ch}}{\mathrm{d}\eta}
\bigg|_{\eta=\eta^{\ast}}
\Bigg/\frac{\mathrm{d}N_{ch}}{\mathrm{d}\eta} \bigg|_{\eta=0},
\end{equation}
where $\mathrm{d}N_{ch}/\mathrm{d}\eta$ is the pseudo-rapidity
distribution of charge particles predicted in Ref.\cite{CGC}. The
transverse momentum of the quark in the medium is reduced to
$p_{T}^{'}=p_{T} - \Delta E$, while the rapidity is kept
unchanged.

Heavy quarks that lose most of their energy due to medium effects
are redistributed according to a thermal
distribution\cite{LinHVQ},
\begin{equation}
\frac{\mathrm{d}N_{\mathrm{thermal}}}{\mathrm{d}m_\mathrm{T}}
\propto m_{\mathrm{T}}~exp(-\frac{m_{\mathrm{T}}}{T}),
\end{equation}
where we choose $T=300~\mathrm{MeV}$\cite{MassiveQuenching}.

\section{Heavy quarks contribution to single muons}
\label{Decay} The hadronization of heavy quarks is assumed to
follow the Peterson fragmentation function\cite{Peterson},
\begin{equation}
D_{h/Q}(z) \propto \frac{1}{z[1-1/z-\epsilon_{Q}/(1-z)]^2},
\end{equation}
where $z=p_{h}/p_{Q}$ is the fraction of heavy quark momentum
carried by the final state hadron. $\epsilon_{Q}$ is a parameter
for the fragmentation, $\epsilon_{c}=0.021$ for charm and
$\epsilon_{b}=0.001$ for beauty are used
here\cite{Benchmark}.
\begin{figure}[tbh]
\centerline{\psfig{file=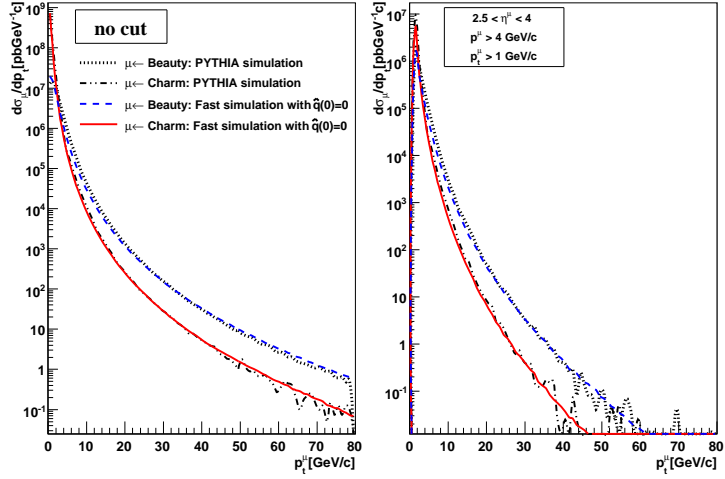,width=0.76\textwidth}}
\vspace*{8pt} \caption{The transverse momentum spectra of single
muons from heavy quarks in pp collisions at
$\sqrt{s_\mathrm{NN}}=5.5~$TeV obtained from the PYTHIA simulation
and the Fast Simulation composed of the HVQMNR program, Peterson
fragmentation and the spectator model. $k_{t}$ broadening and
nuclear shadowing effects are included. The left plot is with no
cut, the right one is within the acceptance of ALICE Muon
Spectrometer. Muons from the decay $B\rightarrow D\rightarrow \mu$
are not taken into account.} \label{Compare}
\end{figure}
\begin{figure}[thb]
\centerline{\psfig{file=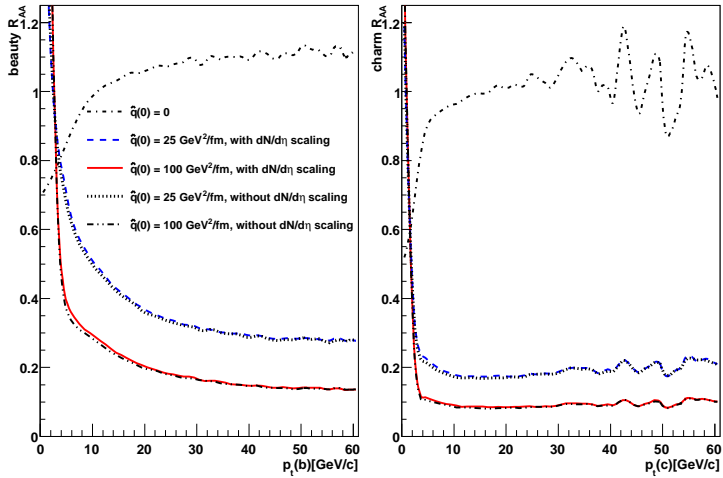,width=0.76\textwidth}}
\vspace*{8pt} \caption{Nuclear Modification factor for heavy
quarks in the central(0-5\%) PbPb collisions at
$\sqrt{s_\mathrm{NN}}=5.5~$TeV. Lines for different $\hat{q}$
values with/without $dN/d\eta$ scaling are given. The left plot is
for beauty and the right one is for charm.}\label{QRaa}
\end{figure}

Heavy mesons produced after the fragmentation decay
semileptonically. In the spectator model, the heavy quark in a
meson is considered to be independent of the light quark and
allowed to decay as a free particle with the simple V-A weak
current\cite{V-A}. Thus, we considered heavy quark
three-body-decay($c\rightarrow\mu+\nu+s$ and
$b\rightarrow\mu+\nu+c$) into muons instead of heavy meson decays
into muons. The decays are calculated in the rest frame of the
heavy quark and then boosted back according to the heavy quark
momentum to obtain the kinematic distributions of
muons\cite{MuonAlice1,MuonAlice2,DimuonEloss}.

\section{Results}
\begin{figure}[thb]
\centerline{\psfig{file=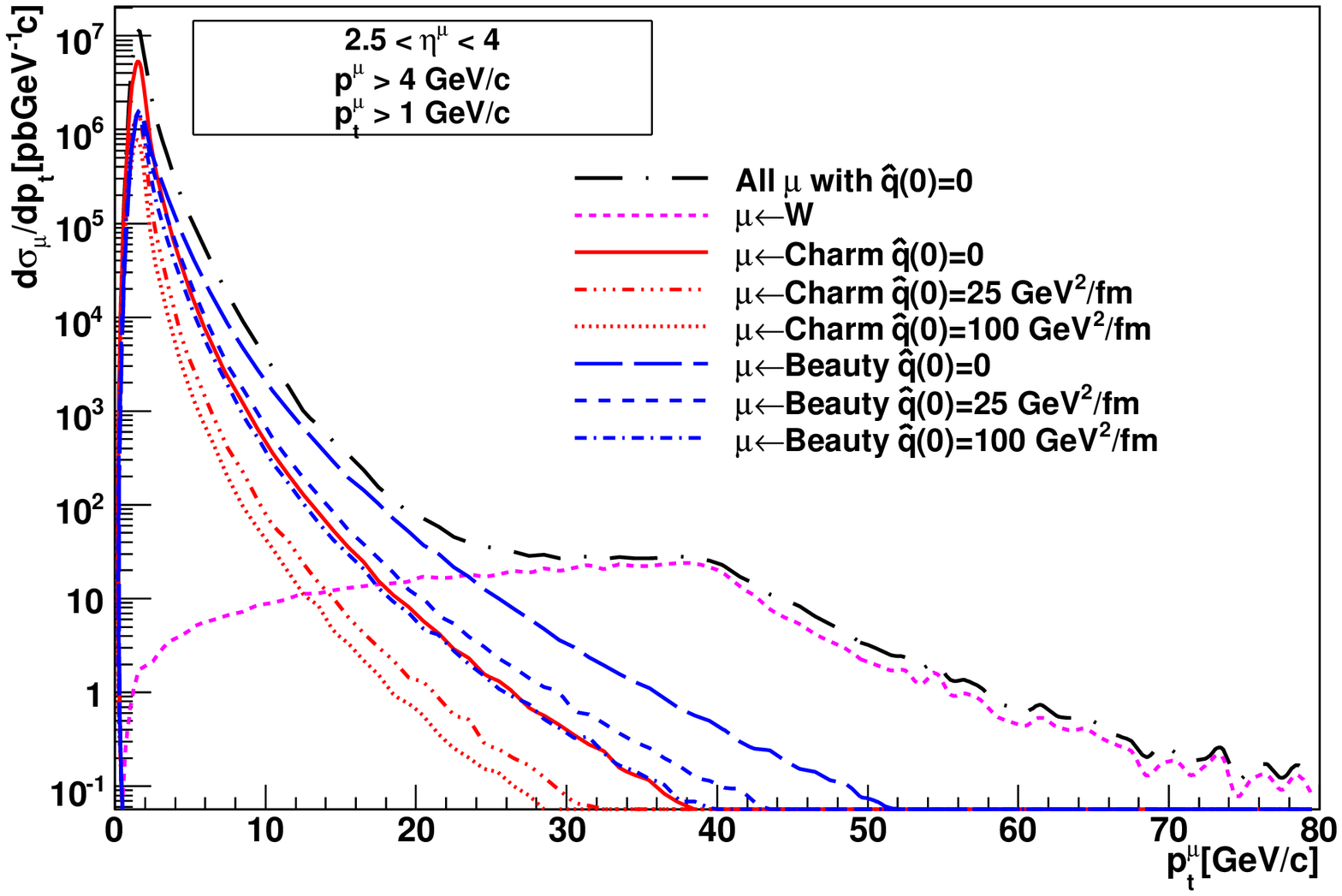,width=0.85\textwidth}}
\vspace*{8pt} \caption{Transverse momentum spectra of single muons
in central(0-5\%) PbPb collisions at
$\sqrt{s_\mathrm{NN}}=5.5~$TeV within the acceptance of ALICE Muon
Spectrometer. Transport coefficients are all applied with
$dN/d\eta$ scaling. The $p_{t}$ spectra of all muons and the muons
from W are obtained from Ref.$^{8}$. The $p_{t}$ spectra of the
muons from heavy quarks are obtained by the Fast Simulation.}
\label{ptMuonAll}
\end{figure}
Fig. \ref{Compare} shows the transverse momentum distribution of
single muons from heavy quarks in pp collisions at
$\sqrt{s_\mathrm{NN}}=5.5~$TeV obtained from PYTHIA\cite{PYTHIA}
simulations(PYTHIA has been tuned to reproduce the kinematic
distribution of heavy quarks produced by the HVQMNR program with
parameter setting described in Section 2) and our Fast Simulation
composed of the HVQMNR program, Peterson fragmentation and the
spectator model(mentioned in the Section \ref{Decay}). $k_{t}$
broadening and nuclear shadowing are included. In Fig.
\ref{Compare} the left plot is without any cuts and the right one
is within the acceptance of the ALICE Muon Spectrometer($2.5 <
\eta^{\mu} < 4,~p^{\mu} > 4~\mathrm{GeV}/c,~p_{t}^{\mu}>1
~\mathrm{GeV}/c$). Muons from the decay $B\rightarrow D\rightarrow
\mu$ are not included. The Fast Simulation gives transverse
momentum distributions of single muons from heavy quarks close to
those obtained from PYTHIA\cite{Zaida}.

To see how much the transport coefficient $dN/d\eta$ scaling
(formula (2)) affects the energy loss of heavy quarks, nuclear
modification factors for different values of $\hat{q}$
with/without $dN/d\eta$ scaling are shown in Fig. \ref{QRaa}. The
left plot and right plot are for beauty and charm, respectively.
From Fig. \ref{QRaa}, we can get that the decreased energy density
in the forward rapidity region makes very small effects on the
energy loss of heavy quarks.


Fig. \ref{ptMuonAll} presents the transverse momentum spectra of
single muons in central(0-5\%) PbPb collisions at
$\sqrt{s_\mathrm{NN}}=5.5~$TeV within the acceptance of ALICE Muon
Spectrometer($2.5 < \eta^{\mu} < 4,~ p^{\mu} > 4~\mathrm{GeV}/c,
~p_{t}^{\mu}>1~\mathrm{GeV}/c$). The $p_{t}$ spectra of all muons
and the muons from W are obtained from Ref.\cite{Zaida}. The
$p_{t}$ spectra of the muons from heavy quarks are obtained by the
Fast Simulation. The values of $\hat{q}(0)$ are set to be 25, 100
GeV$^{2}$/fm\cite{MassiveQuenching,PQM}. As the muons from W
bosons are not effected by the medium formed in the collision,
they can be used as a reference to quantify the effect of heavy
quark energy loss on the single muon production. From Fig.
\ref{ptMuonAll}, we can see that when energy loss is applied to
heavy quarks, the crossing points of the $p_{t}$ spectra of single
muons from heavy quarks and from W are shifted to the lower
$p_{t}$(when $\hat{q}(0)=100~\mathrm{GeV}^{2}$/fm, for beauty, the
crossing point is shifted from 23 to about 17 GeV$/c$, for charm,
it is shifted from 18 to about 12 GeV$/c$).
\section{Conclusions}
We obtain similar $p_{t}$ spectra of single muons from a Fast
Simulation(composed of the HVQMNR program, Peterson fragmentation
and the spectator model) and from PYTHIA generator. Accounting for
the decrease of the medium density in the forward rapidity region,
we assume a $dN/d\eta$ scaling to $\hat{q}$ and we find that the
decreased energy density in the forward rapidity region makes
small effects on the energy loss of heavy quarks. The energy loss
of heavy quarks causes the crossing points in $p_{t}$ spectra of
single muons from heavy quarks and from W bosons shift to lower
$p_{t}$.

\section*{Acknowledgements}
The authors would like to thank F. Antinori and the ALICE group
members in Padova/Legnaro, Italy. And G. Mart\'{\i}nez-Garc\'{\i}a
is gratefully acknowledged for his helpful discussion and
enthusiastic help for the work. This work is supported partly by
the Grant of Museo Storico della Fisica e Centro Studi e Ricerche
'Enrico Fermi', the National Natural Science Foundation of China
under Grant No. 10575044 and Key Grant No.10635020, the Key
Project of the Chinese Ministry of Education No.306022, and partly
by the EU Integrated Infrastructure Initiative Hadron Physics
Project under contract number RII3-CT-2004-506078.

\end{document}